\documentclass[aip,jcp,twocolumn,10pt]{revtex4-1}%endfloats
\usepackage{graphicx}%
\usepackage{dcolumn}
\usepackage{amsmath}   
\usepackage[usenames,dvipsnames]{pstricks}
\usepackage{pst-grad} % For gradients
\usepackage{pst-plot} % For axes

\usepackage{bm}
\usepackage{longtable}

\begin{document}
\title{Solvated dissipative electro-elastic network model
  of hydrated proteins}

\author{Daniel R.\ Martin and Dmitry V.\ Matyushov }\email{dmitrym@asu.edu}
\affiliation{Center for Biological Physics, Arizona State University,
  PO Box 871504, Tempe, AZ 85287-1504 } 
\begin{abstract}
  Elastic netwok models coarse grain proteins into a network of
  residue beads connected by springs. We add dissipative dynamics to
  this mechanical system by applying overdamped Langevin equations of
  motion to normal-mode vibrations of the network. In addition, the
  network is made heterogeneous and softened at the protein surface by
  accounting for hydration of the ionized residues. Solvation changes
  the network Hessian in two ways. Diagonal solvation terms soften the
  spring constants and off-diagonal dipole-dipole terms correlate
  displacements of the ionized residues. The model is used to
  formulate the response functions of the electrostatic potential and
  electric field appearing in theories of redox reactions and
  spectroscopy. We also formulate the dielectric response of the
  protein and find that solvation of the surface ionized residues
  leads to a slow relaxation peak in the dielectric loss spectrum,
  about two orders of magnitude slower than the main peak of protein
  relaxation.  Finally, the solvated network is used to formulate the
  allosteric response of the protein to ion binding. The global
  thermodynamics of ion binding is not strongly affected by the
  network solvation, but it dramatically enhances conformational
  changes in response to placing a charge at the active site of the
  protein.
\end{abstract}
%\pacs{87.14.E-, 87.15.H-, 87.15.kr, 87.10.Pq}
\maketitle

\section{Introduction}
\label{sec:1}
Folding a globular protein in water largely places polar/ionized
residues to its surface, while moving the non-polar residues to its
core. The resulting structure is not unique, and a number of
conformations with close energy minima always exist. Conformational
changes are required for function. They are achieved by either
populating the existing (quasi)stable states (sampling of pre-existing
equilibria\cite{Kern:03,Changeux:2005kx}) or by shifting the existing
minimum-energy conformation to a new configuration minimum upon
perturbation, such as ligand binding (induced fit
mechanism\cite{Zhuravlev:10}).

Conformational transitions involve several types of free energy
penalty. The free energy of elastic deformation relative to the native
structure, involving global shape alteration of the protein, is the
most prominent penalty.\cite{Miyashita:2003ly} This is not the only
long-ranged component of the overall protein's thermodynamics since
electrostatic interactions are also involved in several ways. Changing
the protein conformation alters the interactions between its atomic
charges, but also, to a significant extent, the free energy of
solvation of these charges by hydration water. Water clearly affects
the flexibility of proteins.\cite{Fenimore:04} As a fast, highly polar
subsystem, it follows adiabatically the large-scale protein motions,
continuously stretching and loosening the protein structure by strong
protein-water solvation forces. It lowers the barriers of transitions
between the local minima of the rugged landscape at the energy bottom
of the native basin of attraction,\cite{Thirumalai:10} accelerating
the rate of conformational changes.\cite{Papoian:2004ys} When dried,
proteins stiffen and their relaxation time, as probed by dielectric
spectroscopy, increases by about six orders of
magnitude.\cite{KhodadadiJPCB:08}

The goal of this paper is to develop an efficient computational
algorithm to include hydration in calculations of conformational
flexibility of large protein complexes. Our starting point is to
coarse-grain the protein into an elastic network of beads, a formalism
known as elastic network model
(ENM).\cite{Tirion:96,Atilgan:01,Tama:2001tg} These types of models
aim at calculating the elastic energy of deformation near the
equilibrium structure and the directions of normal-mode displacements
corresponding to the slowest normal-mode vibrations. The typical
coarse-graining is achieved on the scale of a single residue by
replacing it with a single rigid bead. The beads are then connected by
elastic springs physically capturing the connectivity, shape, and
packing of the residues in the folded protein structure.

The ENM coarse-graining of the elastic energy has proven to be very
successful.\cite{Tama:2005oq,Moritsugu:2007ff,Riccardi:09,Romo:2011oq,Lyman:08,hinsen:10766,Miller:2008uq}
Global elastic deformations of a protein are mostly affected by its
shape and mass distribution.\cite{Halle:02,Lu:2005mi,Tama:06}
Electrostatics is another good candidate for coarse-graining. Coulomb
forces are long-ranged and effectively average out the variations of
the local structure. The final outcome for the free energy of
electrostatic interactions is mostly determined by the overall density
and distribution of the protein charge and the dipolar polarization
of the hydration water. This physical reality is addressed by
generalized Born solvation models designing fast computational
algorithms to calculate the free energy of electrostatic
solvation.\cite{Feig:04}

The problem addressed here is two-fold.  First, we want to
re-normalize the elastic network by water's hydration. Given the large
free energy of hydration of the surface residues, the network of beads
is expected to be softer at the interface. We achieve this goal here
by integrating out the dipolar polarization of the hydration water
using formalisms developed in the liquid-state theory of polar
liquids.\cite{DMjcp2:04,DMjcp2:08} The result is an analytical model,
a solvated dissipative electro-elastic network model (sDENM), which
assigns lower force constants to springs attached to ionized
interfacial residues. The second issue is the calculation of the
response functions related to problems affected by the protein
electrostatics. Here, we consider three types of problems: (i)
electrostatic response to a probe charge or dipole placed inside the
protein, (ii) response of the protein to a uniform external field
(dielectric spectroscopy), and (iii) elastic response at a given site
of the protein to altering the charge state of a distant residue
(allosteric action).

\begin{figure}
  \centering
  \includegraphics*[width=5cm]{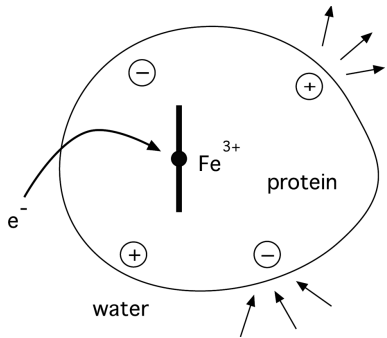}
  \caption{Cartoon displaying the electrostatic perturbation induced
    by a half redox reaction of transferring electron to the heme in
    the protein used as an example. The elastic deformation of the
    protein shifts the ionized surface residues, shown by charges at
    the surface, but also results in adiabatic movements of water
    dipoles solvating them (shown by arrows). }
  \label{fig:1}
\end{figure}

The first type of problems appear in redox reactions involving
proteins\cite{Gray:05} and in optical and IR
spectroscopy\cite{Pal:04,Yang:2010qf} when the position of a spectral
line is affected by the local electric field.  In redox reactions,
electron is transferred by tunneling from an electron donor to an
active site (shown as protein heme in Fig.\ \ref{fig:1}). The dynamics
of this processes, and the activation barrier required to produce
resonant conditions for electron tunneling, can be calculated from the
electrostatic response function $\chi_{\phi}(\omega)$. It arises from
an elastic deformation, shifting the atomic charges of the protein,
caused by transferring the electron, but also from the change in the
positions and orientations of water dipoles hydrating the surface
residues (Fig.\ \ref{fig:1}).  For spectroscopic applications, it is
the dipole moment of the chromophore that is altered by light
absorption. The corresponding response functions $\chi_E(\omega)$ is
the one of the electric field acting on the chromophore dipole and
responsible for spectral solvatochromism.\cite{DMjpca:01}

The second problem addressed here is the dynamic susceptibility of the
protein to a uniform external electric field produced in the
dielectric spectroscopy experiment. Dielectric spectroscopy of
partially wet protein powders has identified a number of generic
relaxation peaks, the assignment of which has been
problematic.\cite{KhodadadiJPCB:08,Khodadadi:2010qf} The
solvent-renormalized network model developed here results in three
relaxation peaks of the protein assigned to fast backbone vibrations
(fastest), the global shape-altering movements (main peak), and the
slow motions of highly solvated charged residues with significant
extent of solvent exposure (slowest).

\begin{figure}
  \centering
  \includegraphics*[width=5cm]{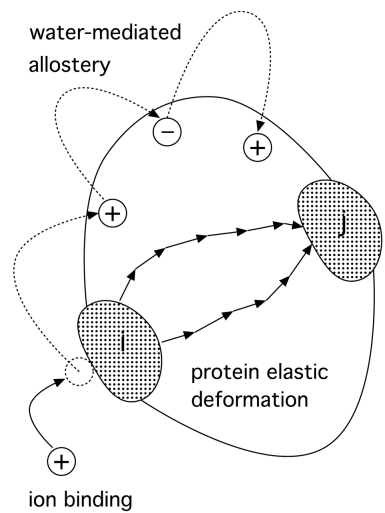}
  \caption{Cartoon showing propagation of piezoelectric perturbation
    caused by binding an ion to residue $i$ and producing a
    displacement of residue $j$. The electric force exerted by the ion
    is propagated throughout the protein as an elastic deformation
    indicated by chains of arrows. In contrast, the displacements of
    ionized surface residues are propagated as water-mediated,
    dipole-correlated surface motions. Ionized surface residues
    combine into a global, correlated net for transmitting signals,
    which does not necessarily require a specific binding site.  }
  \label{fig:2}
\end{figure}

Finally, the last type of problems considered here is the allosteric
response.\cite{Kern:03,Cui:08,Sol:2009uq} It specifies the alteration
in the structure of the protein produced by a perturbation at a
distant site.\cite{Changeux:2005kx} The perturbation can be achieved
by localized ligand binding, often carrying a charge (Fig.\
\ref{fig:2}). Allosteric signaling usually involves oligomeric
proteins, although single-domain proteins also display
allostery.\cite{Volkman:01,Kern:03,Ma:2007ly} Given that two
equilibrium conformations are involved, two equilibrium sets of atomic
coordinates need to be considered for a full description of allosteric
signaling. The barrier to the transition between the two equilibrium
structures is the free energy of the protein elastic deformation,
which can be approximated as crossing of two harmonic elastic
wells.\cite{Miyashita:2003ly,Maragakis:05,Zheng:07,Chu:2007vn,Tripathi:11}
The elastic free energy is quadratic as a function of global
normal-mode displacements, but can change its functional form to a
linear function when localized (cracking) excitations, corresponding
to local unfolding events, are produced.\cite{Miyashita:2003ly}

The lowest elastic barrier is reached along the lowest curvature path
on the free energy surface against deformation, i.e., the lowest
frequency of the elastic vibration. The allosteric pathways are
therefore often associated with the lowest frequencies of harmonic
motions near the two equilibrium structures.\cite{Bahar:05} The
dissipative dynamics of these motions can therefore be explored in the
framework of response functions referring to a single equilibrium
conformation.\cite{Ikeguchi:2005vn} The formulation of such response
functions in the framework of a dissipative electro-elastic network is
our purpose here.

We calculate the dynamics of displacement of a distant residue in
response to changing the charge at the binding site of an allosteric
protein. The main question here is how water modifies the response. We
find that solvation of ionized residues provides a potential mechanism
alternative to the typically anticipated elastic propagation of the
perturbation.

Despite differences in packing and connectivity of residues in
different regions of a folded protein, elastic response tends to be
non-specific, spreading out over the entire volume of the protein
(Fig.\ \ref{fig:2}). In contrast, a net of ionized residues
potentially provides an alternative, surface-bound propagation of the
perturbation by water-mediated allostery. The water-mediated
cross-coupling between the displacements of ionized residue scales as
$r^{-3}$. Therefore, an alteration of the charge at a bindig site can
propagate large distances over the network of surface residues, instead
of, or an addition to, the bulk elastic deformation. The allosteric
response can then be channeled to a site where a conformational change
is required for function.

\section{Model}
\label{sec:2}
The Hamiltonian of the protein hydrated by polar water can be
generally written in the following form
\begin{equation}
  \label{eq:1}
  H = E(\mathbf{R}) - \sum_{i,j} \mathbf{E}_{ij}\cdot \mathbf{m}_j .
\end{equation}
Here, $E(\mathbf{R})$ is the solvent-unperturbed Hamiltonian of the
protein depending on the manifold of atomic coordinates
$\mathbf{R}$. Further, $\mathbf{E}_{ij}$ is the electric field acting
from residue $i$ of the protein on dipole moment $\mathbf{m}_j$ of
water. Both the protein coordinates $\mathbf{R}$ and the water dipoles
$\mathbf{m}_j$ fluctuate with the instantaneous configuration of the
protein-water system; summation over all residues $i=1,\dots,N$ and
all waters $j=1,\dots,N_s$ is taken in Eq.\ \eqref{eq:1}.

Several approximations need to be made in the transition from the
general Hamiltonian in Eq.\ \eqref{eq:1} to a solvated elastic
network. The first approximation is the assumption that an equilibrium
configuration of the protein atomic coordinates is available from the
structural data, and quadratic expansion in atomic displacements can
be done around it. The model thus deals with one conformational state
of the protein only and has nothing to say about transitions between
distinct protein conformations.

The first step in coarse-graining the model is to replace the
collection of protein atomic coordinates with a collection of
beads. We will follow here the standard
approach\cite{Tirion:96,Atilgan:01,Tama:2001tg} of representing each
residue with a single bead, with its position given by the coordinates
of the C$^{\alpha}$ atom. The quadratic expansion of $E(\mathbf{R})$ in
small displacements $\delta r_{i}^{\alpha}=r_i^{\alpha}-
r_{0,i}^{\alpha}$ of individual beads relative to equilibrium
positions $\mathbf{r}_{0,i}$ leads to the relation
\begin{equation}
  \label{eq:2}
    E = (C/2) \sum_{i,j} H_{ij}^{\alpha\beta} \delta
    r_{i}^{\alpha}\delta r_{j}^{\beta}  
\end{equation} 
in which $H_{ij}^{\alpha\beta}$ is a $3N\times 3N$ Hessian matrix and
$C$ is the scaling force constant; $\alpha,\beta$ indicate the
Cartesian projections, and summation over repeated Greek indices is
assumed.

The electrostatic component of the problem is represented by the
standard formulation of atomic force fields. This implies that each
atom of the protein carries the charge $q_{ik}$, where $i=1,\dots,N$
numbers the residues and $k$ represents an atom within residue
$i$. The linear, in the displacements $\delta \mathbf{r}_{i}$,
expansion of the protein-water interaction term in Eq.\ \eqref{eq:1}
results in the following equation
\begin{equation}
  \label{eq:3}
  \delta\mathbf{E}_i\cdot \mathbf{m}_j = \sum_k q_{ik}\delta
  \mathbf{r}_i \cdot \mathbf{T}_{ij}\cdot\mathbf{m}_j ,
\end{equation}
where
$\mathbf{T}_{ij}=-\nabla_i\nabla_j|\mathbf{r}_{0i}-\mathbf{r}_j|^{-1}$
is the dipolar tensor connecting the C$^{\alpha}$ of residue $i$ with
the dipole of water $j$.

In the elastic network constructed here all atoms of a given residue
experience a uniform displacement $\delta\mathbf{r}_i$ from their
equilibrium positions. The librations of the residues are therefore
neglected. This approximation leads to a significant simplification in
Eq.\ \eqref{eq:3} since only charged residues with $\sum_k q_{ik} =
q_i\ne 0$ contribute to the sum. Clearly, uniform displacements of
only charged residues contribute to the creation of the dipole moment
fluctuation $\delta \bm{\mu}_i = q_i\delta \mathbf{r}_i$. We therefore
get for the energy of the protein-water system
\begin{equation}
  \label{eq:5}
  H = (C/2) \sum_{i,j} H_{ij}^{\alpha\beta} \delta
    r_{i}^{\alpha}\delta r_{j}^{\beta} - \sum_i q_i\delta r_i^{\alpha}
    T_{ij}^{\alpha\beta} m_j^{\beta} . 
\end{equation}

We now proceed to calculating the free energy of the hydrated protein
by tracing out the fluctuations of the dipole moments of
water. Adiabatic approximation is assumed at this step, that is we
assume that water is a fast subsystem, equilibrating to each
instantaneous configuration of the network of beads. Since only
global, relatively slow motions of the protein are modeled by the
elastic network, this approximation is expected to be accurate.

Averaging over the configurations of the dipole moments of water
produces partial free energy, i.e., free energy depending on the
manifold of instantaneous displacements $\delta\mathbf{r}_i$. If the
fluctuations of the dipolar polarization field of water are Gaussian,
this free energy is given by the following equation
\begin{equation}
  \label{eq:6}
  F = (C/2) \sum_{ij} \tilde H_{ij}^{\alpha\beta} \delta
  r_i^{\alpha}\delta r_j^{\beta} .
\end{equation}
where the Hessian matrix, renormalized by solvation, becomes
\begin{equation}
  \label{eq:66}
  \tilde H_{ij}^{\alpha\beta} = H_{ij}^{\alpha\beta} -
  C^{-1} \kappa_{ij}^{\alpha\beta} q_i q_j . 
\end{equation}
In Eq.\ \eqref{eq:66}, $\bm{\kappa}_{ij}$ is the rank-2 tensor
representing the dipolar response of the solvent to a dipole
$\delta\bm{\mu}_j$ created at the residue $j$. The dipolar
polarization field created in the solvent in response to this
perturbation then propagates to induce the dipole $\delta\bm{\mu}_i$
at residue $i$. The corresponding free energy cost contributes to
the renormalization of the Hookean force constants for the residues
involved, as represented by the second term in Eq.\ \eqref{eq:66}.

The physical meaning of the solvation terms in Eqs.\ \eqref{eq:6} and
\eqref{eq:66} is quite clear. At $i=j$, the second term in Eq.\
\eqref{eq:66} represents the solvation free energy of the fluctuation
dipole $\delta \bm{\mu}_i$. Correspondingly, $i\ne j$ terms are the
water-mediated couplings of the dipolar fluctuations through the
solvent polarization. Note that direct Coulomb interactions between
$\delta\bm{\mu}_i$ and $\delta\bm{\mu}_j$ are not included in
$\bm{\kappa}_{ij}$. These electrostatic terms propagate the elastic
perturbation through the solvent, by hopping between the ionized
residues, in addition to the direct propagation of elastic forces
through elastic contacts of neighbors in the network (Fig.\
\ref{fig:2}). We note that the standard coarse-grained models of
protein electrostatics, such as generalized Born
models,\cite{Still:90} do not include off-diagonal terms in their
solvation free energy. These terms are however sufficiently
long-ranged, scaling as $r_{ij}^{-3}$ with the distance between the
residues, and they can potentially modify the response of the hydrated
protein to either mechanical or electrostatic perturbation.

\subsection{Polar response of hydration water}
The dipolar susceptibility $\bm{\kappa}_{ij}$ in Eq.\ \eqref{eq:66}
generally requires either liquid-state models of solvation or
electrostatic continuum approaches for its calculation. Here, we start
with the former to introduce a sequence of steps to reduce the full
complexity of polar response to a clear physical picture and a
computationally efficient algorithm.

The susceptibility $\bm{\kappa}_{ij}$ is given by the convolution of
the dipolar tensors $\mathbf{T}_i= \mathbf{T}(\mathbf{r}_{0i}-\mathbf{r})$,
representing the electric field of the dipole $\delta\bm{\mu}_i$ at the
point $\mathbf{r}$ in water, with the spacial correlation function of
the dipolar fluctuations of water interfacing the protein. It can be
conveniently represented by the convolution of inverted space
$\mathbf{k}$-integrals\cite{DMjcp2:04,DMcp:06,DMjcp2:08} 
\begin{equation}
  \label{eq:7}
  \bm{\kappa}_{ij} = \mathbf{\tilde T}_i(\mathbf{k})*
      \bm{\chi}(\mathbf{k},\mathbf{k}') * 
      \mathbf{T}_j(\mathbf{k}'). 
\end{equation}
Here, the asterisks between tensors represent tensor contraction over
common indices and integration over common $\mathbf{k}$-variables.
Further, the response function $\bm{\chi}(\mathbf{k},\mathbf{k}')$
depends on two wave-vectors to reflect the inhomogeneous nature of the
problem caused by the presence of the protein in solution. Finally,
$\mathbf{\tilde T}_i(\mathbf{k})$ is the Fourier transform of the
dipolar tensor taken over the volume $\Omega$ occupied by water
\begin{equation}
  \label{eq:4}
  \mathbf{\tilde T}_i(\mathbf{k}) = \int_{\Omega}\mathbf{T}(\mathbf{r} -
  \mathbf{r}_{0i}) \theta(\mathbf{r}) e^{i\mathbf{k}\cdot \mathbf{r}} d\mathbf{r}.
\end{equation}

As we have shown elsewhere,\cite{DMjcp2:04} the nonlocal part of
$\bm{\chi}(\mathbf{k},\mathbf{k}')$ is mostly due to transverse
polarization fluctuations, given by the component of the dipolar
polarization perpendicular to the unit vector $\mathbf{\hat
  k}=\mathbf{k}/k$.\cite{Madden:84,Kivelson:89} This transverse
response is in fact fairly small for most solvation
problems\cite{DMjcp1:04} (e.g., the Born solvation energy is entirely
longitudinal) and will be neglected here. This approximation
eliminates the dependence on the second wave-vector with the
result\cite{DMjcp2:04}
\begin{equation}
  \label{eq:8}
  \bm{\chi}(\mathbf{k},\mathbf{k}') = \mathbf{\hat k}\mathbf{\hat k}
  \chi_s^L(k) (2\pi)^3\delta(\mathbf{k}-\mathbf{k}') .
\end{equation}
Here, $\chi_s^L(k)$ is the longitudinal dipolar susceptibility of the
homogeneous liquid depending on the scalar magnitude $k$ only. It is
typically given as a product of the density of dipoles in the liquid
$y$ and the longitudinal structure factor
$S^L(k)$:\cite{Kivelson:89,Skaf:95} $\chi_s^L(k)=(3y/4\pi)
S^L(k)$. The dipolar density parameter $y=(4\pi/9)\beta \rho m^2$ is
defined by the liquid number density $\rho$ and molecular dipole $m$;
$\beta=1/(k_{\text{B}}T)$ is the inverse temperature.

With the form of the response function given by Eq.\ \eqref{eq:8}, the 
convolution in Eq.\ \eqref{eq:7} is reduced to a 3D integral. While
this problem is numerically tractable,\cite{DMcp:06,DMjcp2:08} the 
number of integrals to be evaluated is $\sim N_i^2/2$, where $N_i$ is
the number of ionized residues. This is still a numerically intense
computation, and simplifications are desired. 

We will further simplify the problem by modeling the calculation of the 
dipolar tensor of a given residue in Eq.\ \eqref{eq:4}. The full
calculation of the Fourier transform requires numerical integration over the
volume outside the typically complex shape of the protein.\cite{DMjcp2:08} 
To avoid this computationally extensive step, the concept of
relative accessible surface area\cite{Hasel:88} will be employed
here. Specifically, the volume integral in Eq.\ \eqref{eq:4} will be
replaced with the integral outside the sphere of radius $s$,
representing the average distance of the closest approach of the water
molecules to the residue, and scaled with the
fraction of the surface area $\alpha_i$ exposed to the solvent   
\begin{equation}
  \label{eq:9}
  \mathbf{\tilde T}_i (\mathbf{k}) = - 4\pi \mathbf{D} \alpha_i 
          \frac{j_1(ks)}{ks} e^{i\mathbf{k}\cdot\mathbf{r}_{0i}} .
\end{equation}
In this equation, $\mathbf{D}=3\mathbf{\hat k}\mathbf{\hat k} -
\mathbf{1}$, $j_n(x)$ is the spherical Bessel function, and
$\alpha_i$ is the ratio of the solvent exposed area $a_i$ to the
overall surface area of the residue 
\begin{equation}
  \label{eq:10}
  \alpha_i = a_i /(4\pi s^2 ).
\end{equation}

The reduction of Eq.\ \eqref{eq:9} yields an analytical solution for
the solvent response function
\begin{equation}
  \label{eq:11}
\begin{split}
  \kappa_{ij}^{\alpha\beta} =  \frac{4y\alpha_i\alpha_j}{3} 
   &\bigg[\frac{a(s,r)}{s^3}\delta_{\alpha\beta}\delta_{ij}\\
                          & -  (1-\delta_{ij})
     b(s,r) T_{ij}^{\alpha\beta}
    \bigg]. 
\end{split}
\end{equation}
Here, indices are dropped for brevity in $r=r_{ij}$ and
$T_{ij}^{\alpha\beta}$ is the direct-space dipolar tensor connecting
beads $i$ and $j$. The first summand in the brackets in Eq.\
\eqref{eq:11} represents the solvation free energy of the dipole at a
charged bead and the second term represents the dipole-dipole
interaction between the two charged beads.

The coefficients $a(s,r)$ and $b(s,r)$ in Eq.\ \eqref{eq:11} are
obtained as one-dimensional integrals including the longitudinal
structure factor of the liquid $S^L(k)$ to account for non-local
correlations between the solvent dipoles. These integrls are listed
and calculated in the Appendix.  We show there that the dependence on
$s$ and $r$ can be lifted in these functions and they in fact are well
represented by constants, $a(s,r)=S^L(0)A$, $b(s,r)=S^L(0)$. Here,
$S^L(0)=(3y)^{-1}(1-\epsilon_s^{-1})$ represents the longitudinal dielectric
response of a homogeneous polar liquid with the dielectric constant
$\epsilon_s$. We finally get for the solvation tensor
\begin{equation}
  \label{eq:13}
   \kappa_{ij}^{\alpha\beta} = \frac{4\alpha_i\alpha_j}{9}\left(1-
     \frac{1}{\epsilon_s} \right)
   \left[\frac{A}{s^3}\delta_{\alpha\beta}\delta_{ij} - 
     (1-\delta_{ij})T_{ij}^{\alpha\beta} \right] .
\end{equation}
The constant $A=3.54$ is calculated in the Appendix assuming $s=4.4$
\AA\ for the closest-approach distance between the center of a surface
amino acid and the oxygen of water. The result is not strongly
affected by the choice of $s$. Note, however, that $A$ absorbs the
thermodynamic state of the solvent into it and will change with its
thermodynamic state (temperature, pressure, etc) through the
corresponding alterations of the polarization structure factor. All
calculations and MD simulations presented here refer to the
temperature of 300 K and ambient pressure.

\subsection{Dissipative elastic network}
The free energy of the hydrated protein in Eq.\ \eqref{eq:6} is a
quadratic form in residues' displacements $\delta\mathbf{r}_i$. It can
be used to calculate the response to external perturbations or
equilibrium variances once the network Hessian $H_{ij}^{\alpha\beta}$
has been specified. We will use here the Hookean springs Hamiltonian
suggested by Tirion.\cite{Tirion:96} This potential, $E(\mathbf{R}) =
(C/2) \sum_{ij} D_{ij} (r_{ij} -r_{0,ij})^2$, describes the elongation
$r_{ij}=|\mathbf{r}_i-\mathbf{r}_j|$ between nodes $i$ and $j$ in the
network characterized by one universal force constant $C$
($r_{0,ij}=|\mathbf{r}_{0,i}-\mathbf{r}_{0,j}|$). In this potential,
$D_{ij}$ is the connectivity matrix. Its value is set to unity when
$r_{ij}$ is within the cutoff distance $r_c$ and is set to zero
otherwise. In addition, $D_{ij}=\varepsilon>1$ for covalently bound
neighbors. This scaling accounts for a stronger bonding of residues
along the backbone, and is known to better model the vibrational
density of states of the protein.\cite{Ming:2005fj} Finally, the
renormalization of the network by solvation of and electrostatic
interactions between ionized residues will follow Eqs.\ \eqref{eq:6}
and \eqref{eq:66}.

The equations of motion for the network beads need to be specified in
order to calculate the time-dependent response
functions.\cite{Hansen:03} The elastic network obviously lacks
dissipative dynamics typical for soft condense phases. Alternatives to
purely mechanical equations of motions can be sought in terms of
Langevin dynamics of individual
beads.\cite{Miller:2008uq,Essiz:2009fk} Introducing dissipation at the
level of individual beads is not necessarily an obvious
choice,\cite{Soheilifard:2011pi} and we have previously opted to
introduce dissipation to normal modes $\mathbf{q}_m$ diagonalizing the
network Hessian.\cite{DMpb:12} The equation of motion for such
overdamped dynamics is \cite{Hansen:03}
\begin{equation}
  \label{eq:14}
  \int_0^{t} \zeta(t-t') \mathbf{\dot{q}}_m(t') dt' + \lambda_m
  \mathbf{q}_m = \mathbf{F}(t) + \mathbf{R}(t),
\end{equation}
where $\zeta(t-t')$ is a memory function,
$\mathbf{F}(t)=\mathbf{F}_{\omega}e^{i\omega t}$ is an external
oscillating force, and $\mathbf{R}(t)$ is a randomly fluctuating
force. The latter satisfies the generalized fluctuation-dissipation
relations \cite{Kubo:66,Zwanzig:01}
\begin{equation}
  \label{eq:15}
  \langle \mathbf{R}(t) \rangle =0, \quad \langle\mathbf{R}(t)\cdot
  \mathbf{R}(0)\rangle = k_{\text{B}}T\zeta(t) .
\end{equation}
The eigenvalues $\lambda_m$ of normal modes $q_m$ in this equation are
obtained by diagonalizing the Hessian in Eq.\ \eqref{eq:66}. They are
therefore affected by solvation softening the interface. We indeed
observe a shift of the vibrational density of states to softer modes
when the network is solvated.

Applying Laplace-Fourier transform \cite{Hansen:03} to Eq.\
\eqref{eq:14} results in the displacement response function for the
collective mode $\mathbf{q}_m$.  It is given as a scalar
function connecting the average displacement to the external
field,\cite{Kubo:66} $\langle \mathbf{q}_m(\omega)\rangle =
\chi_m(\omega)\mathbf{F}_{\omega}$, where $\langle
\mathbf{q}_m(t)\rangle =\langle \mathbf{q}_m(\omega)\rangle e^{i\omega
  t}$. From Eq.\ \eqref{eq:14}, one gets
\begin{equation}
  \label{eq:16}
  \chi_m(\omega) = \left[ i\omega\tilde \zeta(\omega) + \lambda_m\right]^{-1} ,
\end{equation}
where $\tilde \zeta(\omega)$ is the Laplace-Fourier transforms of the
friction kernel $\zeta(t)$. The entire set of $3N$ eigenvalues
$\lambda_m$ is produced by diagonalizing the Hessian with the unitary
matrix $\mathbf{U}$. The inclusion of all eigenvalues of the Hessian
results in the response function of the bead displacements
\begin{equation}
  \label{eq:17}
   \chi_{ij}^{\alpha\beta} (\omega) = C^{-1} \sum_{m}
   U_{mi}^{\gamma\alpha} \chi_m(\omega) U_{mj}^{\gamma\beta} .
\end{equation}

The distance- and self-correlation functions of protein residues
typically show two characteristic relaxation times of overdamped
motion and, correspondingly, two Debye peaks in their loss
spectra. Therefore, following the prescription of our previous
work,\cite{DMpb:12} we use a two-Debye form of $\chi_m(\omega)$, which
two characteristic friction coefficients, $\zeta_l$ and $\zeta_h$
\begin{equation}
  \label{eq:27}
  \chi_m(\omega) = \frac{a}{i\omega \zeta_h + \lambda_m} +
  \frac{1-a}{i\omega\zeta_l +\lambda_m}, 
\end{equation}
where the amplitude $a$ specifies the relative weight of each
relaxation component. 

\subsection{Electrostatic response functions}
The network response function $\bm{\chi}_{ij}(\omega)$ in Eq.\
\eqref{eq:17} describes the displacement of residue $i$ induced by a
weak oscillating force applied to residue $j$. Since the residue
displacement uniformly moves all of it atomic charges, this linear
susceptibility can be used to build electrostatic response functions
of either electrostatic potential or electric field at a given
location within the protein.\cite{DMpb:12}

Assume that an oscillatory probe charge $q_0(t)=q_{\omega}e^{i\omega
  t}$ is placed at some location $\mathbf{r}_0$ within the protein.  This
charge will act on residue $i$ with the force
$-q_0(t)\mathbf{E}_{0i}$, where $\mathbf{E}_{0i}$ is the electric field
produced at $\mathbf{r}_0$ by all charges of residue $i$ at their
equilibrium positions 
\begin{equation}
  \label{eq:21}
  \mathbf{E}_{0i} = \sum_k 
 \frac{q_{ik} (\mathbf{r}_0 - \mathbf{r}_{ik})}{|\mathbf{r}_{ik}-\mathbf{r}_0|^3}
\end{equation}
Here, $q_{ik}$ are the atomic charges of residue $i$ with the
equilibrium coordinates $\mathbf{r}_{ik}$.

The force acting on residue $i$ will propagate through the elastic
network to residue $j$ according to the response function
$\bm{\chi}_{ij}(\omega)$. The displacement of that residue will in
turn produce an alteration of the electrostatic potential of the
protein at $\mathbf{r}_0$. After summing over all residues in the
network, one arrives at the frequency-dependent susceptibility of the
electrostatic potential
\begin{equation}
  \label{eq:18}
  \chi_{\phi}(\omega) = - \sum_{i,j}
  E_{0j}^{\alpha}\chi_{ij}^{\alpha\beta}(\omega) E_{0i}^{\beta} . 
\end{equation} 
This susceptibility determines the alteration of the electrostatic
potential produced by the charges of the protein matrix $\delta
\phi_0(\omega)$ at the position $\mathbf{r}_0$ of the probe charge
$q_{\omega}$: $\delta\phi_0(\omega) =
\chi_{\phi}(\omega)q_{\omega}$.

Similarly, one can define the electric field alteration
$\delta\mathbf{E}_0$ produced by the protein matrix at the point
$\mathbf{r}_0$ in response to placing an oscillating probe dipole
$\bm{\mu}_{\omega}$ at that point.  This frequency-dependent
susceptibility is based on convoluting $\bm{\chi}_{ij}(\omega)$ with
the dipolar tensors
$\mathbf{T}_{ik}=\mathbf{T}(\mathbf{r}_{0}-\mathbf{r}_{ik})$
connecting the residue charge $q_{ik}$, located at $\mathbf{r}_{ik}$,
to the position of the probe dipole at $\mathbf{r}_0$. The result is
\begin{equation}
  \label{eq:19}
  \chi_E^{\alpha\beta}(\omega)  = \sum_{i,j,k,l} q_{ik} T_{ik}^{\alpha\gamma}
  \chi_{ij}^{\gamma\delta}(\omega) T_{jl}^{\delta\beta} q_{jl} .
\end{equation} 
Here, as above, summation runs over the repeated Greek indices
denoting Cartesian projections of the corresponding tensors. The
difference in signs in Eqs.\ \eqref{eq:18} and \eqref{eq:19} comes
from the fact that the free energy invested into the creation of the
potential alteration is $(1/2)q_{\omega}\delta\phi_0(\omega)$, while
for the dipole one has
$-(1/2)\bm{\mu}_{\omega}\cdot\delta\bm{E}_0(\omega)$.

\subsection{Dielectric response}
When a uniform oscillatory external field
$\mathbf{E}_0(t)=\mathbf{E}_{\omega}e^{i\omega t}$ is applied to a
protein, it induces the dipole moment $\delta \mathbf{M}(\omega)=\sum_j
\delta\bm{\mu}_j(\omega)$. Since only movements of the charged residues produce
non-zero dipoles,  $\delta \mathbf{M}(\omega)=\sum_j
q_j \delta\mathbf{r}_j(\omega)$. Substituting the network
displacements susceptibility, one arrives at the relation
\begin{equation}
  \label{eq:24}
  \delta M^{\alpha}(\omega) = \sum_{i,j} q_i q_j
  \chi_{ij}^{\alpha\beta}(\omega) E_{\omega}^{\beta} .
\end{equation}
Assuming that the external field is along the $z$-axis of the
laboratory frame, one gets for the dipolar susceptibility
\begin{equation}
  \label{eq:25}
  \chi_M (\omega) = \sum_{ij} q_i q_j \chi_{ij}^{zz}(\omega) = 
 (1/3) \sum_{ij} q_i q_j \chi_{ij}^{\alpha\alpha}(\omega) .
\end{equation}

This dipolar susceptibility refers to the dipole moment induced at a
single protein molecule. It can be used to calculate the
complex-valued dielectric constant $\epsilon_p(\omega)$ of a protein
sample (either powder or polycrystal) by applying the standard
derivation of the theory of dielectrics.\cite{Scaife:98} The result is
\begin{equation}
  \label{eq:26}
  \frac{(\epsilon_p(\omega)-1)(2\epsilon_(\omega)+1)}{9\epsilon_p(\omega)}
  = \frac{4\pi}{3}\rho_p \chi_M(\omega),
\end{equation}
where $\rho_p$ is the number density of the protein molecules in the
material.

\begin{figure}
  \centering
  \includegraphics*[width=7cm]{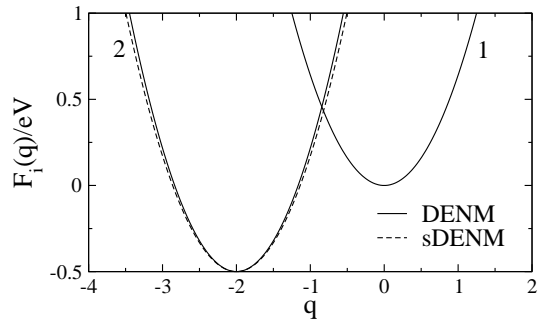}
  \caption{Free energy surfaces representing the free energy penalty
    (reversible work) of changing the charge at a bindig site of an
    allosteric protein. The calculations are done for attaching
    carbamoylphospate ($q_{02}=-2$) to the bacterial enhancer-binding
    protein NtrC (Fig.\ \ref{fig:8}). The curves refer to DENM in the
    unphosphorylated state ($q_{01}=0$) and to DENM and sDENM in the
    phosphorylated state ($q_{02}=-2$). The elastic network is defined
    with $k_{\text{B}}T/C=0.75$ \AA$^2$ and $\epsilon=125$; the cutoff
    radius is 15 \AA. The free energy of binding $\Delta F$ is unknown
    and was set at $-0.5$ eV for the purpose of illustration.  }
  \label{fig:3}
\end{figure}

\subsection{Allosteric response}
As an example of the application of the formalism of response
functions to the allosteric response of a protein, we will consider
the displacement $\delta r_i^{\alpha}(\omega)$ of residue $i$ in
response to binding a charge $q(t)= q_{\omega}e^{i\omega t}$ at
position $\mathbf{r}_0$. The response is therefore effectively of
piezoelectric type,\cite{Landau8} creating deformation at a distant
site in response to electric stimulus.

Binding of an ion causes both global and local perturbations of the
protein. From the global perspective, it changes the free energy of
the entire protein by the free energy of binding $\Delta F$ and, in
addition, exerts Coulomb forces acting on all charges of the
protein. The global perspective can be studied, as is typically done
in electrostatics,\cite{Landau8} by asking what is the free energy
cost $F_i(q)$ of transferring a small probe charge $q$ to the binding
site, where $i=1,2$ labels the two stable conformations of the
protein, ion-free and ion-bound. The electrostatic potential
susceptibility [Eq.\ \eqref{eq:18}] addresses this question. The free
energy cost is obviously
\begin{equation}
  \label{eq:28}
  F_i(q) = \frac{1}{2}\chi_{\phi}^{(i)}(0)(q-q_{i0})^2 + F_{i0}, 
\end{equation}
where $q_{10}=0$, $F_{10}=0$ and $q_{20}=q_i$ is the charge of the
binding ion and $F_{20}=\Delta F$ is the binding free energy. We have
also added the dependence of the response function on the protein
state since susceptibility $\chi_{\phi}^{(i)}(0)$ can be sensitive
to structural changes of the protein.

The amount of transferred charge $q$ can be viewed as the reaction
coordinate for the global free energy cost of binding the partial
charge $q$. The crossing of the free energy surfaces,
$F_1(q^{\dag})=F_2(q^{\dag})$, will define the transition state and the
corresponding free energy barrier. This picture allows for a ``Marcus
inverted region behavior'',\cite{MarcusSutin} i.e.\ there is an
optimal binding free energy minimizing the free energy barrier.  The
activation barrier starts to grow when $\Delta F$ falls below the
optimal value (inverted region). In addition, because the curvatures
of the two parabolas may differ, there is a scenario in which no
crossing in the inverted region occurs, i.e.\ the activation barrier
becomes infinite and the reaction is not allowed. Note, however, that
the curvatures of two surfaces calculated for the NtrC protein studied
below are nearly identical (Fig.\ \ref{fig:3}). There is also little
sensitivity of the overall free energy functions to solvation of the
surface residues (compare DENM and sDENM calculations). This lack of
global sensitivity is in stark contrast with a strong effect of
solvation on individual residue displacements, as we show below.

Most of the interest in the field is driven not by the global
thermodynamics of binding, but by the need to understand biological
function caused by it.\cite{Cui:08,Sol:2009uq} A typical problem is to
calculate the displacement of a distant residue in response to
binding. We will approach this question, as above, by considering an
oscillatory charge placed at $\mathbf{r}_0$. This charge will interact
with each residue $j$ by the electric field $\mathbf{E}_{0j}$ given by
Eq.\ \eqref{eq:21}. That interaction creates the force acting on each
bead in the network, $\delta F_{j}^{\alpha}(\omega) = -
E_{j0}^{\alpha} q_{\omega}$.

In the linear response approximation,\cite{Hansen:03} the average
displacement of residue $i$ is given by summing up the forces produced
by the charge $q_{\omega}$ at all residues of the network with their
response function $\chi_{ij}^{\alpha\beta}(\omega)$ propagating the
force at $j$ into a displacement at $i$
\begin{equation}
  \label{eq:22}
  \langle \delta r_i^{\alpha}(\omega) \rangle = - \sum_j
  \chi_{ij}^{\alpha\beta} (\omega) E_{j0}^{\beta} q_{\omega} . 
\end{equation}
In the calculations below we will present the scalar displacement 
\begin{equation}
  \label{eq:23}
  \delta r_i(\omega) = \left[ \langle \delta r_i'^{\alpha}(\omega) \rangle 
\langle \delta r_i'^{\alpha}(\omega) \rangle \right]^{1/2}, 
\end{equation}
where $\delta r_i'^{\alpha}(\omega)$ is the real part of the
complex-valued displacement and summation over repeated Greek indices
is performed.

\begin{figure}
  \centering
  \includegraphics*[width=7cm]{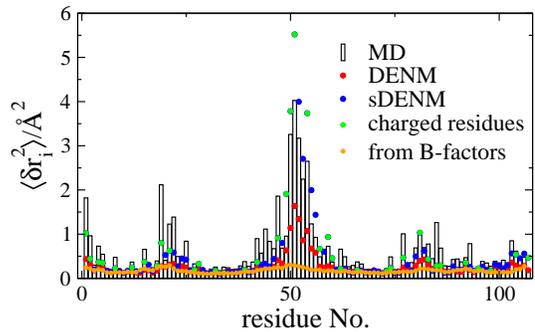}
  \caption{Mean-square displacements of cytochrome B562 (cytB)
    C$^{\alpha}$'s from MD simulations, DENM, and sDENM
    calculations. The network parameters in the DENM/sDENM
    calculations are: $k_{\text{B}}T/C=0.75$ \AA$^2$, $\epsilon=125$,
    cut-off radius is 15 \AA. The solvent accessible surface for the
    loop residues labeled in Fig.\ \ref{fig:5} is scaled down to 45
    \AA$^2$. }
  \label{fig:4}
\end{figure}

The frequency $\omega$ of the oscillatory charge might effectively
represent the time-scale of charge binding, such as the frequency of
binding/unbinding events, typically occurring on the nanosecond
time-scale for small electrolyte ions.\cite{Zhuravlev:10} The limit
$\omega=0$ in this formalism will represent stationary, i.e.,
adiabatically slow binding.

\section{Results}
Before presenting the results of specific calculations, we start with
some crude estimates of the effect of solvation of charged residues on
the properties of the elastic network. Our previous calculations of
electrostatic properties of redox proteins were done with the elastic
spring constant of $C=0.6$ kcal/(mol \AA$^2$), consistent with other
estimates in the literature.\cite{Tirion:96,Atilgan:01} Given this
force constant, one can estimate the effect of solvation on the
network Hessian. We consider the diagonal element in Eq.\
\eqref{eq:15},
\begin{equation}
  \label{eq:12}
  \tilde H_{ii}^{\alpha\alpha} = 2 -
  4A\alpha_i^2q^2/(3Cs^3)(1-\epsilon_s^{-1}) .
\end{equation}
With $q=e$, $s=4.4$ \AA, $\epsilon_s=78$, and $A=3.54$, the second
term becomes $30\alpha_i^2$. This estimate suggests that any
singly-charged residue exposed to water to more than a quater of its
surface will have a negative elastic constant with its non-covalent
neighbors and will be held in the equilibrium position only by
covalent bonds within the network. Such ionized residue would lose
mechanical stability and dissolve in water if not held in place by its
covalent neighbors. It is clear that solvation makes a major effect on
the elastic response of charged residues.

\begin{figure}
  \centering
  \includegraphics*[width=8cm]{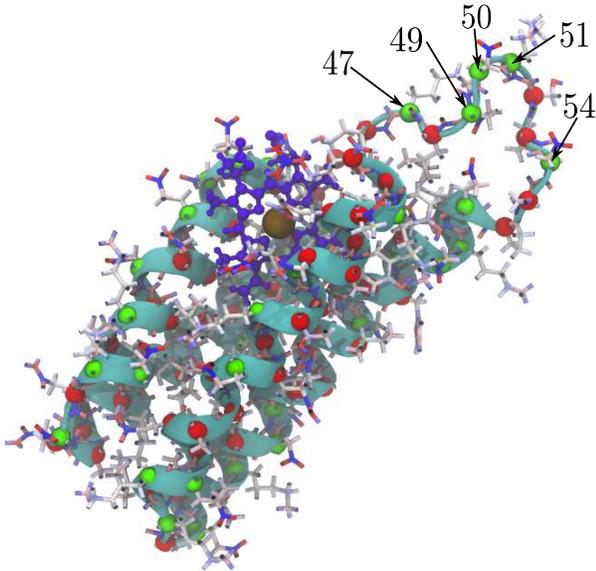}
  \caption{Cartoon of cytochrome B562 (cytB) showing the positions of
    C$^{\alpha}$ (spheres) colored by residue charge: charged (green)
    and uncharged (red). The charged residues marked green (also green
    points in Fig.\ \ref{fig:4}) are also required to have $\alpha_i$
    greater than 0.16, used as a threshold number. The remaining
    C$^{\alpha}$ are marked red. The side chain atoms are colored by
    charge: negative (red), positive (blue), and neutral (white).  The
    heme iron is colored brown while the remaining atoms of the heme
    are blue.  Numbers label the unstable residues in the loop for
    which the water-exposed surface was scaled down to 45 \AA$^2$ in
    order to maintain the network stability. }
  \label{fig:5}
\end{figure}

\subsection{Residue displacements and electrostatic response}
The standard approach of experimental verification and
parameterization of elastic protein networks is to compare the
root-mean-square displacements (rmsd's) of residues with experiment or
Molecular Dynamics (MD) simulations. Crystallographic B-factors are
often used,\cite{Bahar:10aa} but those are of limited
value.\cite{Halle:02} It was noted that reported B-factors are
dominated by rigid-body motions of the proteins in the
crystal.\cite{Soheilifard:08} In addition, there is a clear mismatch
between the reported rmsd's of proteins in crystals and in their
flexibility in solution, as is illustrated in Fig.\ \ref{fig:4}
comparing rmsd's from B-factor of C$^{\alpha}$'s with their rmsd's
found from MD. The MD simulations were done for hydrated cytochrome
B562 (cytB, PDB entry 256B, Fig.\ \ref{fig:5}) as described
elsewhere.\cite{DMjpcb:11,DMpb:12}

The mismatch between both the B-factors and the standard ENM as
compared to MD is particularly notable for the flexible loop (residues
46 to 55) containing several ionized residues (Figs.\ \ref{fig:4} and
\ref{fig:5}). This region is clearly not restricted to a single
configuration in solution and instead wanders through a number of
semi-stable conformations. The network, required to reside in a single
conformation, is expected to lose stability because of this and
similar segments.  The standard ENM clearly avoids this instability by
over-restricting the flexible residues. In contrast, when
renormalization by solvation is introduced in sDENM, the network loses
stability, as expected, due to solvation of the loop residues labeled
in Fig.\ \ref{fig:5}. Since calculations cannot be performed with an
unstable network, we have artificially restricted the network by
scaling down the solvent-accessible area of the residues labeled in
Fig.\ \ref{fig:5} from the values calculated with VMD\cite{VMD} (in
the range 120--150 \AA$^2$) to 45 \AA$^2$. This rescaling prevents
mechanical instability of the network, but preserves the physical
reality of a flexible loop, as is seen from the corresponding rmsd's
in Fig.\ \ref{fig:4}.

\begin{figure}
  \centering
  \includegraphics*[width=7cm]{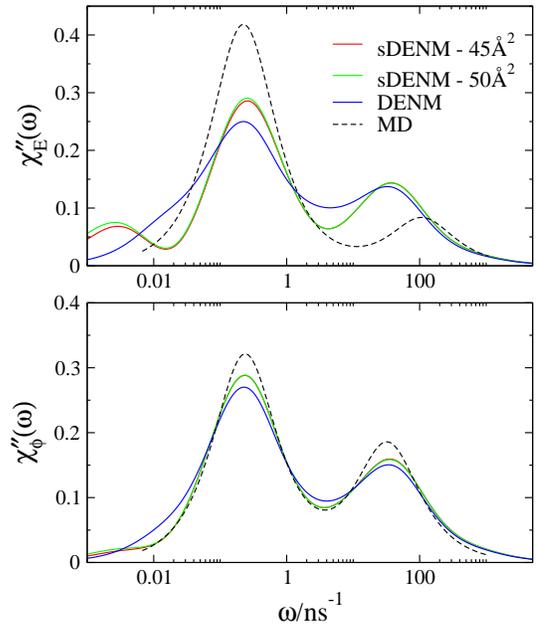}
  \caption{Loss spectra $\chi_E''(\omega)/\chi_E'(0)$ and
    $\chi''_{\phi}(\omega)/\chi_{\phi}'(0)$ for cytB. Compared are MD,
    DENM, and sDENM calculations. To show the sensitivity of sDENM
    calculations to solvation of the loop residues in cytB, the
    results of choosing the solvent-accesible area of $a_i=45$ \AA$^2$
    and of $a_i=50$ \AA$^2$ are shown. The two-Debye relaxation
    parameters are $\zeta_l = 30 $ ns, $\zeta_h=0.006\zeta_l$, and
    $a=0.35$ [Eq.\ \eqref{eq:27}]. The elastic network is defined with
    $k_{\text{B}}T/C=0.75$ \AA$^2$, $\epsilon=125$, and the cutoff
    radius of 15 \AA. }
  \label{fig:6}
\end{figure}

Figure \ref{fig:6} shows the results of the calculations (cytB) for
the electrostatic potential susceptibility $\chi_{\phi}(\omega)$ and
the tensor contraction $\chi_E(\omega)=\chi_E^{\alpha\alpha}(\omega)$
for the electric field susceptibility. The effect of solvating surface
residues is less pronounced for these susceptibilities, in particular
for the more long-ranged electrostatic potential. A slow relaxation
component, not resolved on the length of the MD trajectory, appears
for the electric field susceptibility.  This slow component arises from
much slower motions of highly solvated residues in the network, also
seen in the dielectric response of the protein.

\subsection{Dielectric susceptibility of the protein}
\begin{figure}
  \centering
  \includegraphics*[width=7cm]{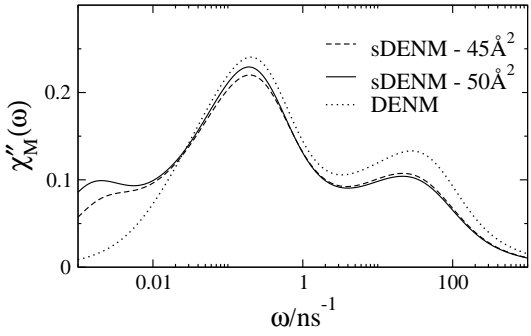}
\caption{$\chi_M''(\omega)/ \chi'_M(0)$ for cytB. The parameters of the
  network are the same as in Fig.\ \ref{fig:6}. }
\label{fig:7}
\end{figure}

The calculated imaginary part of the dielectric susceptibility (loss
function) $\chi''_M(\omega)$ in shown in Fig.\ \ref{fig:7}. Similarly
to the case of $\chi_E''(\omega)$, it clearly shows the emergence of a
slow peak, about two orders of magnitude slower than the main peak.
The slow component comes from the hydrated residues with high exposure
to water. This is clearly seen from the sensitivity of the slow peak
to the solvent-accessible surface assigned to the residues of the
loop. We need to note that the network completely neglects librations
of polar residues, focusing only on the polarization fluctuations
produced by translational motions of the charged residues. An
additional dielectric intensity might therefore come from the
components missing from the model.

Experimentally, partially hydrated protein powders show three
relaxation processes at low temperatures, which merge into two
processes at ambient temperature.\cite{KhodadadiJPCB:08} The fastest
and the slowest processes disappear when the protein is dried. The
relaxation time of the main peak from dielectric measurements matches
well the relaxation time from neutron scattering experiments, in which
the protein and hydration water signals can be separated by
deuteration. The main peak is therefore assigned to global protein
motions.\cite{KhodadadiJPCB:08} In this regard, the main peak in Fig.\
\ref{fig:7} can be tentatively aligned with the main peak of
dielectric measurements.

The slowest observed peak,\cite{KhodadadiJPCB:08} about two orders of
magnitude slower than the main peak, has been hard to interpret by
experimental means. Its strong dependence on the level of hydration,
however, suggests that it should be linked to the protein. Indeed, our
calculations give clear evidence that slow relaxation is related to
overdamped motions of highly solvated ionized residues. The two orders
of magnitude ratio of the slow and main relaxation times is in
qualitative agreement with the dielectric measurements. Finally, the
peak disappears when hydration of ionized residues is removed,
which is analogous to drying the sample in experiment.  The assignment
of the slow peak should also emphasize the involvement of water in the
relaxation process. Since water is fast and follows adiabatically the
protein motions, ionized residues move by dragging hydration waters
with them.

\begin{figure}
  \centering
  \includegraphics*[width=7cm]{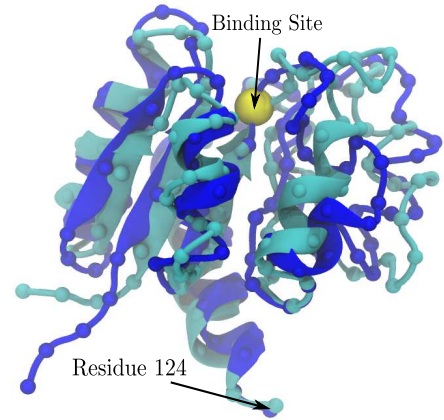}
  \caption{Superimposed structures of the bacterial enhancer-binding
    protein NtrC in dephosphorylated (light blue,PDB entry 1DC7) and
    phosphorylated (dark blue, PDB entry 1DC8)
    states.\cite{Kern:99,Volkman:01} The NMR structure was
    determined\cite{Kern:99} with carbamoylphospate binding to
    Asp54. The displacement of Glu124 in response to a probe charge at
    Asp54 is shown in Fig.\ \ref{fig:9}.}
  \label{fig:8}
\end{figure}

\subsection{Allosteric response} 
The calculations of the allosteric response to ion binding have been
done for a single-domain signaling protein NtrC. The structures of
this protein have been resolved\cite{Kern:99} both in unphosphorylated
(denoted as NtrC) and in phosphorylated (denoted as P-NtrC)
states. The latter state is short-lived, and it was maintained in
solution at a large excess of the phosphordonor carbamoylphosphate
carrying the charge of $q_{02}=-2$. The addition of this charge to
Asp54 active site (Fig.\ \ref{fig:8}) creates a Coulomb force acting
on the neighboring atomic charges, such that each residue $j$
experiences the force $-q_{02}\mathbf{E}_{0j}$. The perturbing force
produced by ion binding is therefore fairly nonlocal, in contrast to a
common assumption,\cite{Ikeguchi:2005vn} and the calculation of the
response requires full account of this fact.

\begin{figure}
  \centering
  \includegraphics*[width=7cm]{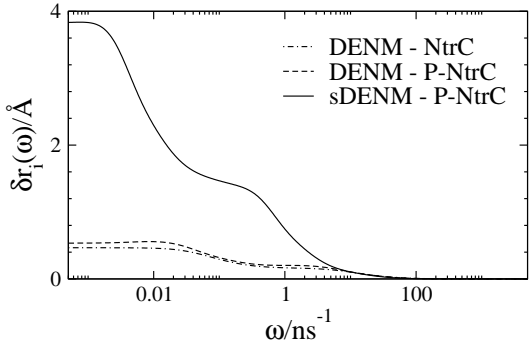}
  \caption{Frequency-dependent allosteric displacement [Eq.\
    \eqref{eq:23}] for residue $i=124$ (Glu) of NtrC in the
    unphosphorylated (NtrC) and phosphorylated (P-NtrC) states. The
    results of calculations within DENM and sDENM are compared as
    shown in the plot. }
  \label{fig:9}
\end{figure}

The frequency-dependent displacement of residue $i$ in Eqs.\
\eqref{eq:22} and \eqref{eq:23} sums up all Coulomb forces acting on
residues $j$ from the active site labeled as ``0''. The
unphosphorylated (NtrC) state of the protein is very mobile, with
several loops continuously changing their conformation on the $\mu$s
to ms time-scale. These motions mostly disappear in a more compact
phosphorylated state.\cite{Volkman:01} Not surprisingly, we have found
that sDENM is rather unstable and only DENM calculations could be done
on the NtrC state. Therefore, sDENM calculations were done only on the
P-NtrC state. The results for $\delta r_{124}(\omega)$ in both states
are shown in Fig.\ \ref{fig:9}. As expected, the inclusion of
solvation in sDENM greatly enhances the displacement magnitude. The
frequency dependence is also noteworthy. It implies the existence of
$\mu$s motions of the protein responding to the charge
perturbation. \cite{Volkman:01} If the frequency of binding/unbinding
events exceeds this frequency, the protein does not have the ability
to respond to the perturbation and this subset of motions dynamically
freezes. As a result, the displacement diminishes.

\begin{figure}
    \centering
    \includegraphics*[width=7cm]{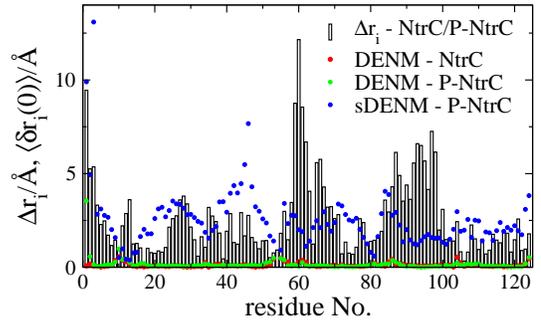}
    \caption{Displacements $\Delta r_i$ between two equilibrium
      structures (NtrC and P-NtrC) of the NtrC protein. $\delta
      r_i(0)$ shows the displacement of residue $i$ in response to
      placing a unitary probe charge $q_{\omega}=1$ at the position of
      C$^{\alpha}$ of the binding site (Asp54, Fig.\ \ref{fig:8}).  }
    \label{fig:10}
\end{figure}

Figure \ref{fig:10} emphasizes a strong effect of solvation on ion
binding allostery presented in Fig.\ \ref{fig:9} by showing
zero-frequency displacements of all residues in the NtrC protein. The
calculations have been done in DENM for NtrC and P-NtrC and in sDENM
for P-NtrC. The results are compared with $\Delta r_i$ displacements
of C$^{\alpha}$ between the two structures (Fig.\ \ref{fig:8}). It is
clear that only by including solvation within sDENM does the
displacements of residues in a given state reach the magnitudes
comparable with the overall displacement amplitudes $\Delta r_i$. It
appears that, while solvation does not strongly affect the global
energetics of ion binding (Fig.\ \ref{fig:3}), it critically affects
the allosteric amplification of ion binding through conformational
transitions of individual residues.

\section{Summary}
Folded proteins have to maintain structural stability. At the same
time many functions of enzymes and motor proteins involve large-scale
domain movements in response to binding and release of ligands. This
requirement makes one suggest that some stability needs to be
sacrificed to allow amplification of a small perturbation into a large
response. The question is what are the structural motifs that allow
amplification without compromising the global stability.  A related
issue is the length of correlations, since long-ranged correlations
are required for allosteric action at the distance.

The first obvious target to address the problem is elasticity. The
protein is densely packed and any force perturbation propagates
through its body as in a glass material. The elastic deformation
spreads, however, through the elastic body and does not accommodate
for a directed, specific action.  Elasticity can capture motions of
relatively rigid domains linked by flexible
hinges,\cite{Ikeguchi:2005vn} but to a lesser extent the allostery of
monomeric, single-domain systems.

One can alternatively turn attention to hydration water\cite{Ball:08}
as a possible medium for transferring signals. Water can store
significant energy in the form of dipolar polarization and large
entropy in its hydrogen-bond network, but it is also a highly
non-specific medium. The required specificity might therefore reside
at the protein-water interface combining large energies stored in
hydration with regulation achieved through identities of the surface
residues.

The surface charges, and to some extent dipoles, carry large solvation
free energies and are strongly correlated in their motions, with
water-mediated correlations decaying as $r^{-3}$. Because of
non-locality of correlations, an ensemble of ionized surface residues
forms a strongly correlated net (or, in a sense, merged multiple
pathways\cite{Sol:2009uq}) enveloping the entire protein. Given that
multiple binding sites are typically involved in protein function,
allostery might be designed not by building a specific site and
attaching strings (``communication
pathways''\cite{Datta:2008fk,Zhuravlev:10}) to it, but by puling on
the net wherever one finds a ``knot''. Some knots might be more
important than the others from the perspective of biological
function. The non-locality of this net excludes the possibility of
well-defined pathways, they must be achieved by more specific
interactions involving either no-polar
residues\cite{Bruschweiler:2009ys} or chains of hydrogen
bonds.\cite{Datta:2008fk}

A network of ionized, hydrated surface residues is a general property
of all hydrated proteins, affecting a number of observable
properties. The formalism of solvated dissipative electro-elastic
network captures this reality and projects it on a number of
susceptibilities describing the response to a particular type of external
perturbation of a given experiment. A number of observables, such as
rmsd's, dielectric response, electrostatic susceptibilities, and
allosteric response are affected by solvation of the surface
residues. The general outcome is that elastic motions of the residues
become significantly heterogeneous, with softening achieved at the
sites carrying charges. The interfacial heterogeneity does not
dramatically affect the global thermodynamics of the protein or the
thermodynamics of ion binding, but is critical for local responses to
external perturbations. While global motions of the protein altering
its shape occur on the nanosecond time-scale, $\mu$s
motions\cite{Volkman:01,Kneller:2012zr} are assigned to portions of
the protein with highly hydrated ionized residues.

\acknowledgments This research was supported by the National Science
Foundation (MCB-1157788). CPU time was provided by the National
Science Foundation through TeraGrid resources (TG-MCB080116N).

\appendix
\section{Solvation Integrals}
\label{appA}
The solvation integrals in Eq.\ \eqref{eq:11} are given by the
following one-dimensional $k$-space integrals involving the
longitudinal structure factor $S^L(k)$ of the homogeneous solvent
\begin{equation}
  \label{eq:1A}
  \begin{split}
a(s,r) & = \delta_{r,0} + \frac{6s}{\pi} \int_0^{\infty} j_1(sk)^2
j_0(rk)(S^L(k)-1) dk, \\
b(s,r)& = \theta(r-2s) \\
      &+ \frac{6s}{\pi} \left(\frac{r}{s}\right)^3 
\int_0^{\infty} j_1(sk)^2j_2(rk)(S^L(k)-1) dk . 
  \end{split}
\end{equation}
Here, $j_n(x)$ is a spherical Bessel function, $s$ denotes the
effective radius of a residue, and $r$ is the distance between the
centers of two beads in the network. Further, $r=0$ corresponds to one
bead and that configuration is represented by the Kronecker delta
$\delta_{r,0}$, which is equal to unity when $r=0$. Since dipolar
structure factors satisfy the asymptote $S^L(k)\rightarrow 1$ at $k
\rightarrow \infty$, this limit is separated from the numerical
integral and is given by the first summand in each equation.

\begin{figure}
  \centering
  \includegraphics*[width=6cm]{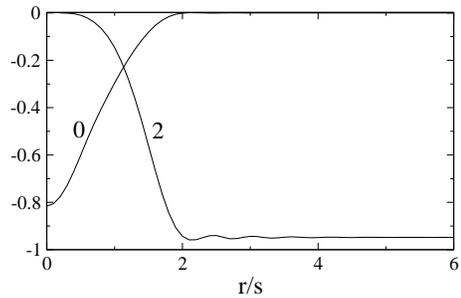}
  \caption{Integrals entering Eq.\ \eqref{eq:1A} calculated with the
    longitudinal structure factor of TIP3P water.\cite{DMjcp2:08} The
    labeling in the plot point to the integration with $j_0(rk)$ in
    (first equation in Eq.\ \eqref{eq:1A}, labeled as ``0'') and with
    $j_2(rk)$ (second equation in Eq.\ \eqref{eq:1A}, labeled as
    ``2''). The integrals are calculated as the function of $r/s$ with
    fixed $s=4.4$ \AA.  }
  \label{fig:11}
\end{figure}

The dipolar structure factors of polar liquids have been intensively
studied in the past.\cite{Skaf:95} Analytical models from liquid-state
theories also exist.\cite{Hansen:03} Several studies reported
structure factors of force-field water
models.\cite{Bopp:96,Omelyan:98} This function is accesible only from
simulations since there is no known experimental technique giving
access to it.

For the purpose of estimating the integrals in Eq.\ \eqref{eq:1A} we
have taken the longitudinal structure factor of TIP3P water calculated
from MD simulations.\cite{DMjcp2:08} A simple parameterization of this
function is available\cite{DMjcp1:04,DMjcp2:08} based on the solution
of mean-spherical closure for dipolar hard spheres.\cite{Hansen:03}
The results of this integration are shown in Fig.\
\ref{fig:11}.  As is seen the first integral involving
$j_0(rk)$ quickly goes to zero when reaching the distance $r/s\simeq
2$. Since $r$ is either zero, for one-bead solvation, or greater than
$2s$, for different beads, only $r=0$ needs to be considered for this
function. We therefore put $a(s,r)=A(s)S^L(0)$, where $A(s =
4.4\mathrm{\AA})=3.54$ is numerically calculated. The value $s=4.4$
\AA\ is the sum of the average radius of 3 \AA\ assigned to a residue
and 1.4 \AA\ for the radius of waters.

The situation is just the opposite for the second integral. It is zero
at $r=0$ and reaches the value $S^L(0)-1$ at $r/s=2$. This latter
result implies that the continuum limit approximation $S^L(k) =
S^L(0)$ applies in this case. We therefore put
$b(s,r)=S^L(0)\theta(r-2s)$.

The overall result of these calculations, incorporating solvent
dipolar correlations through the longitudinal structure factor, is
quite clear. The solvation energy, at $r=0$, is renormalized by the
factor $A$ from the dielectric continuum limit $S^L(k) = S^L(0)$. This
renormalization effectively reduces the cavity radius for dipolar
solvation from the distance of the closest approach of water to the
residue $s$ to $s/A^{1/3}$. This is consistent with the common
observation that the effective cavity radius should fall between $s$
and the van der Waals radius of the solute $s-\sigma_s/2$ ($\sigma_s$
is the water diameter). On the other hand, the dipolar water-mediated
coupling between distant residues is well described by the continuum
limit of the solvent dipolar response, and that fact is reflected in
the constancy of $b(s,r)$.

\bibliography{chem_abbr,dielectric,dm,statmech,protein,liquids,solvation,dynamics,glass,elastic,simulations,surface,bioet,et,nano,photosynthNew,enm,bioenergy,nih,ir}

\end{document}